\documentclass[]{spie}  %>>> use for US letter paper
\usepackage{amsmath,amsfonts,amssymb}
\usepackage{graphicx}
\usepackage[colorlinks=true, allcolors=blue]{hyperref}

\usepackage[export]{adjustbox}
\usepackage{multirow}
\usepackage{subfig}
\usepackage{hhline}
\usepackage{xkeyval,xcolor}% http://ctan.org/pkg/{xkeyval,xcolor}
\makeatletter
\newlength{\sfp@hseplen}\newlength{\sfp@vseplen}
\define@cmdkey{subfigpos}[sfp@]{pos}[ul]{}% \sfp@pos
\define@cmdkey{subfigpos}[sfp@]{font}[\small]{}% \sfp@font
\define@cmdkey{subfigpos}[sfp@]{vsep}[5pt]{\setlength{\sfp@vseplen}{\sfp@vsep}}% \sfp@vsep
\define@cmdkey{subfigpos}[sfp@]{hsep}[5pt]{\setlength{\sfp@hseplen}{\sfp@hsep}}% \sfp@hsep
\newcommand{\subfigimg}[3][,]{%
  \setkeys{Gin,subfigpos}{pos,font,vsep,hsep,#1}% Set (default) keys
  \setbox1=\hbox{\includegraphics{#3}}% Store image in box
  \ifnum\pdfstrcmp{\sfp@pos}{ul}=0% UPPER LEFT placement of subfig label
    \leavevmode\rlap{\usebox1}% Print image
    \rlap{\hspace*{\sfp@hsep}\raisebox{\dimexpr\ht1-\sfp@vsep}{\sfp@font{#2}}}% Print label
    \phantom{\usebox1}% Insert appropriate spacing
  \else\ifnum\pdfstrcmp{\sfp@pos}{ur}=0% UPPER RIGHT placement of subfig label
    \leavevmode\usebox1% Print image
    \llap{\raisebox{\dimexpr\ht1-\sfp@vsep}{\sfp@font{#2}}\hspace*{\sfp@hsep}}% Print label
  \else\ifnum\pdfstrcmp{\sfp@pos}{lr}=0% LOWER RIGHT placement of subfig label
    \leavevmode\usebox1% Print image
    \llap{\raisebox{\sfp@vsep}{\sfp@font{#2}}\hspace*{\sfp@hsep}}% Print label
  \else% Assume LOWER LEFT placement of subfig label
    \leavevmode\rlap{\usebox1}% Print image
    \rlap{\hspace*{\sfp@hseplen}\raisebox{\sfp@vsep}{\sfp@font{#2}}}% Print label
    \phantom{\usebox1}% Insert appropriate spacing
  \fi\fi\fi
}
\makeatother
\title{Improving Small Lesion Segmentation in CT Scans using Intensity Distribution Supervision:\\Application to Small Bowel Carcinoid Tumor}

\author[a]{Seung Yeon Shin}
\author[a]{Thomas C. Shen}
\author[b]{Stephen A. Wank}
\author[a]{Ronald M. Summers}
\affil[a]{Imaging Biomarkers and Computer-Aided Diagnosis Laboratory\\Radiology and Imaging Sciences, Clinical Center, National Institutes of Health, Bethesda, MD, USA}
\affil[b]{Digestive Disease Branch, National Institute of Diabetes and Digestive and Kidney Diseases, National Institutes of Health, Bethesda, MD, USA}

\authorinfo{Further author information: Send correspondence to Seung Yeon Shin (seungyeon.shin@nih.gov) or Ronald Summers (rms@nih.gov)}

\begin{document} 
\maketitle

\begin{abstract}

Finding small lesions is very challenging due to lack of noticeable features, severe class imbalance, as well as the size itself. One approach to improve small lesion segmentation is to reduce the region of interest and inspect it at a higher sensitivity rather than performing it for the entire region. It is usually implemented as sequential or joint segmentation of organ and lesion, which requires additional supervision on organ segmentation. Instead, we propose to utilize an intensity distribution of a target lesion at no additional labeling cost to effectively separate regions where the lesions are possibly located from the background. It is incorporated into network training as an auxiliary task. We applied the proposed method to segmentation of small bowel carcinoid tumors in CT scans. We observed improvements for all metrics (33.5\% $\rightarrow$ 38.2\%, 41.3\% $\rightarrow$ 47.8\%, 30.0\% $\rightarrow$ 35.9\% for the global, per case, and per tumor Dice scores, respectively.) compared to the baseline method, which proves the validity of our idea. Our method can be one option for explicitly incorporating intensity distribution information of a target in network training.
 
\end{abstract}

% Include a list of keywords after the abstract

\section{INTRODUCTION}\label{sec:intro}
%1.5p
% small lesion identification (in CT scans)
Identifying small lesions from a computed tomography (CT) scan is clinically relevant since it can lead to immediate treatment if needed or trigger surveillance~\cite{cai21}. Despite its necessity, it is more difficult than for larger lesions since small lesions are likely to be less noticeable due to the relative lack of imaging features such as shape and texture as well as the size itself.  

% small lesion segm
Especially in training a segmentation model for small lesions, it also entails a higher degree of class imbalance, which makes achieving a reasonable performance difficult. This issue can be mitigated by performing segmentation of an organ-of-interest, which could contain the target lesions, sequentially or jointly with lesion segmentation~\cite{kamble20,tang20}. Lung segmentation can benefit lung nodule segmentation by restricting the region of interest~\cite{kamble20}. In the work by Tang et al.~\cite{tang20}, liver and liver tumor segmentation is performed jointly using the same network to utilize the correlation between them. Features learned for liver segmentation would be relevant to liver tumor segmentation since the tumors reside within it.

% clinical setting
In the work by Ayalew et al.~\cite{ayalew21}, a significantly lower segmentation accuracy was achieved for segmenting tumors from an entire CT volume than conducting it from the identified liver, which may imply the need of restricting the region of interest for better small lesion segmentation. However, training the organ segmentation together with the target lesion segmentation requires additional ground-truth (GT) segmentations of the organ. Indeed, a clinician is interested in finding and marking abnormalities, but not an entire organ.

% intensity distribution (HU) supervision
Intensity values in CT scans (Hounsfield units) carry important information and can provide clues as to which tissue a given voxel belongs to~\cite{buzug11}. An intensity distribution of a target lesion can be used to effectively separate regions where the lesions are possibly located from the other. It can be achieved by investigating intensity values within available GT segmentations of lesions, or can be provided as prior information.

\begin{figure}[t]
	\centering
	\begin{minipage}{1\textwidth}
	    \subfloat{\includegraphics[width = 0.1\textwidth]{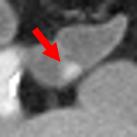}}
	    \subfloat{\includegraphics[width = 0.1\textwidth]{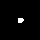}}
        \subfloat{\includegraphics[width = 0.1\textwidth]{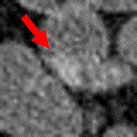}}
	    \subfloat{\includegraphics[width = 0.1\textwidth]{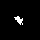}}
	    \subfloat{\includegraphics[width = 0.1\textwidth]{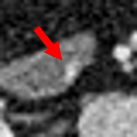}}
	    \subfloat{\includegraphics[width = 0.1\textwidth]{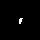}}
	    \subfloat{\includegraphics[width = 0.1\textwidth]{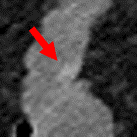}}
	    \subfloat{\includegraphics[width = 0.1\textwidth]{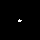}}
	    \subfloat{\includegraphics[width = 0.1\textwidth]{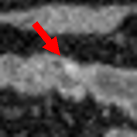}}
	    \subfloat{\includegraphics[width = 0.1\textwidth]{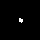}}
    \end{minipage}
	\caption{Examples of small bowel carcinoid tumors in our dataset. In each image, the tumor is pointed by the red arrow. Their corresponding ground-truth (GT) segmentation is also presented.}
	\label{fig:car_tumor_ex}
\end{figure}

% carcinoid tumor (segm)
We segment small bowel carcinoid tumors from CT scans in this paper. To the best of our knowledge, this is the first work that uses only CT scans for this task. In the work by Carlsen et al.\cite{carlsen22}, PET (positron emission tomography) / CT scans are used to segment tumors. Carcinoid tumor is a type of neuroendocrine tumor. Despite being rare (small bowel neoplasms including carcinoid tumors account for $0.5\%$ of all cancers in the United States~\cite{jasti20}), there has been an increase in the diagnosed incidence of carcinoid tumors for the past several decades. While more than half ($50-71.4\%$) of carcinoid tumors develop within the gastrointestinal tract, they are found predominantly in the small bowel ($24-44\%$)~\cite{hughes16}. Figure~\ref{fig:car_tumor_ex} shows example carcinoid tumors in the small bowel. They are submucosal in location and often less than a centimeter in size~\cite{hughes16}. Like the aforementioned organ/lesion segmentation cases, performing small bowel segmentation together could help segmentation of small bowel carcinoid tumors. However, this is very costly due to the difficulty of labeling the small bowel~\cite{shin20,shin21,shin22_spie,shin22_isbi,shin22_miccai}.

% proposed method
Instead, we use the intensity distribution of small bowel carcinoid tumors to compute a probability of being tumor for each voxel. This soft label is provided to our segmentation model as an auxiliary task so that it can inform the model about our region of interest, which could contain carcinoid tumors, based on the intensity. It is a soft and possibly disconnected surrogate of the organ segmentation, which requires no additional annotation cost.

\section{METHOD}\label{sec:method}
%2.5p
\subsection{Dataset}

24 preoperative CT scans of 24 unique patients who underwent surgery and were found to have at least one carcinoid tumor within the small bowel were collected at our institution. They are all intravenous and oral contrast-enhanced abdominal CT scans. For each patient, either arterial phase ($n$=18) or venous phase ($n$=6) scans were selectively used according to the relevant description in the corresponding radiology report. All scans were acquired with oral administration of Volumen using 0.5, 1, or 2 mm slice thickness. They were cropped manually along the z-axis to include from the diaphragm through the pelvis. Then, they were resampled to have isotropic voxels of 1x1x1 mm$^3$.

\begin{figure}[t]
	\centering
	\begin{minipage}{1\textwidth}
	    \subfloat[]{\includegraphics[width = 0.33\textwidth]{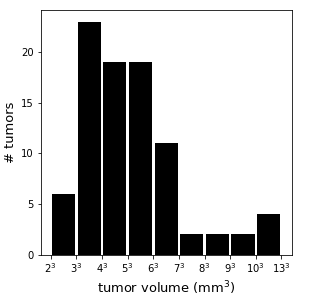}}
	    \subfloat[]{\includegraphics[width = 0.66\textwidth]{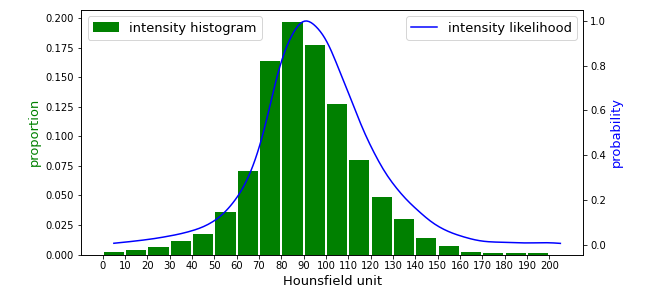}}
    \end{minipage}
	\caption{(a) Tumor volume distribution in our dataset. (b) Tumor intensity distribution (green) and a likelihood function (blue) built from it. They have different scales, so the left or right y-axis should be read for each. Refer to the text for the explanation of the likelihood function.}
	\label{fig:stat}
\end{figure}

GT segmentation of tumors was drawn in CT scans, using ``Segment Editor" module in 3DSlicer~\cite{fedorov12}, by referring to the corresponding radiology report as well as an available 18F-DOPA PET scan of each patient. This resulted in 88 annotated tumors in total. The tumor volume was 8 mm$^3$ (approximately 2 mm diameter) at minimum and 1960 mm$^3$ (approximately 16 mm diameter) at maximum. Figure~\ref{fig:stat} (a) shows the distribution of tumor sizes in our dataset. 

\subsection{Intensity Distribution Supervision}

Figure~\ref{fig:stat} (b) shows the distribution of tumor intensities, which was computed by aggregating intensity values within the GT tumor segmentations in our dataset. Images that have been smoothed with anisotropic diffusion~\cite{black98} are used. Since the histogram is discontinuous around bin boundaries, to make a smooth function from it, kernel density estimation~\cite{parzen62} is performed using Gaussian kernels with automatic bandwidth determination. Finally, it is rescaled to have the maximum value of $1$. The function in Figure~\ref{fig:stat} (b) is the constructed intensity likelihood model. It enables faster calculation of the likelihood for a large number of voxels than using the histogram. We note that the explained procedure requires no additional labeling effort. While it could be less precise, the likelihood model can be provided also from a user as prior information.

\begin{figure}[t]
	\centering
	\includegraphics[width = 1\textwidth]{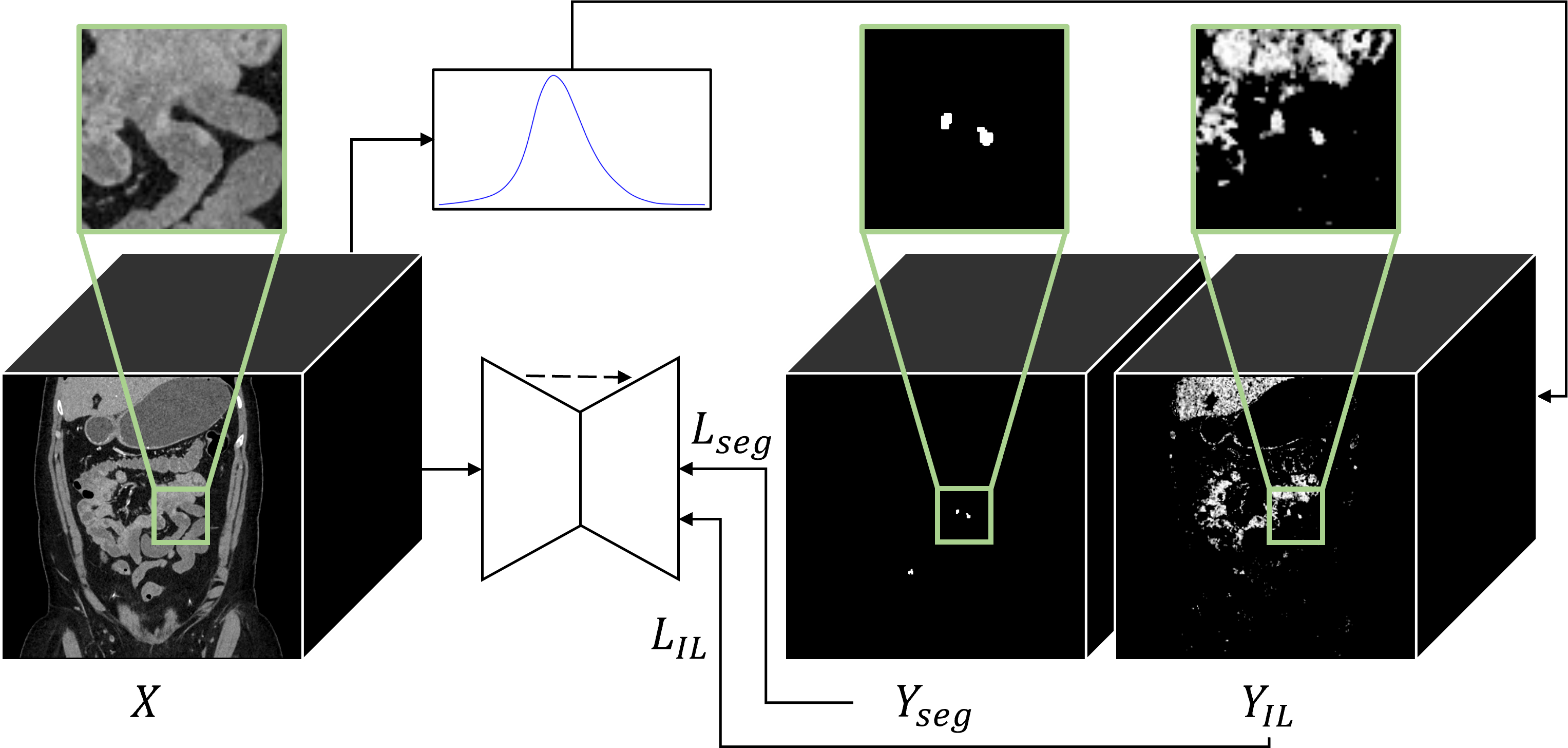}
	\caption{Visualization of data involved in training. An example input volume $X$ and corresponding GT labels $Y_{seg}$, $Y_{IL}$ for training are presented. $Y_{seg}$ and $Y_{IL}$ are GT tumor segmentation and GT intensity likelihood, respectively. $Y_{IL}$ is generated from the input volume $X$ using the likelihood function of Figure~\ref{fig:stat} (b), at no additional labeling cost. The network architecture is based on the structure of the 3D U-Net~\cite{cicek16}. The dotted line represents skip connections. The network predicts two outputs, namely, tumor segmentation and tumor intensity likelihood. They are compared against GT labels $Y_{seg}$ and $Y_{IL}$ to compute their respective losses $L_{seg}$ and $L_{IL}$.}
	\label{fig:data_in_out}
\end{figure}

Figure~\ref{fig:data_in_out} visualizes the network and data involved in training. Given an input volume $X$, a corresponding intensity likelihood volume $Y_{IL}$ can be made using the prepared function, where each voxel value represents a probability of being tumor according to the intensity. Then, we make use of it to augment our tumor segmentation network. Similar to the network used by Tang et al.~\cite{tang20} for joint liver and liver tumor segmentation, we use a network with two output channels in the proposed method. One channel predicts tumor segmentation. The second channel predicts the tumor intensity likelihood in place of organ segmentation. The generated likelihood volume $Y_{IL}$ is used as supervision for this channel. This \emph{soft} label is intended to inform the network about our region of interest, which could contain carcinoid tumors especially in terms of intensity, as \emph{hard} organ segmentation supervision does in the joint organ/lesion segmentation case~\cite{tang20}.

\subsection{Training}

Two loss terms are involved in our network training (Figure~\ref{fig:data_in_out}), which are $L_{seg}$ for tumor segmentation and $L_{IL}$ for tumor intensity likelihood prediction. We use the generalized Dice loss~\cite{sudre17} for $L_{seg}$. For $L_{IL}$, a cross-entropy loss is used to measure the dissimilarity between GT and prediction of the tumor intensity likelihood. Finally, the overall loss function for training the proposed network is:

\begin{equation}
    \label{eq:loss_tatal}
    \begin{aligned}
	L = L_{seg} + \lambda L_{IL}
	\end{aligned}
\end{equation}
where $\lambda$ is the relative weight for $L_{IL}$.

\subsection{Evaluation Details}

Our network architecture is based on that of the 3D U-Net~\cite{cicek16}, but has fewer channels, which are \{8, 16, 32, 64\}. While the final inference layer has 1x1x1 kernels, all the other convolution layers have 3x3x3 kernels. Group normalization~\cite{wu18} is used between each convolution layer and non-linearity function. It is implemented using PyTorch 1.8.2.

We used an NVIDIA Tesla V100 32GB GPU to conduct experiments. For training, sub-volumes of size 224x224x224 sampled from the original volumes are used to fit in the memory. Each sub-volume is guaranteed to have at least one tumor. The mini-batch size was set as $1$. We used the AdamW optimizer~\cite{loshchilov19} and a weight decay of $5 \times 10^{-4}$. Based on the grid search, $3 \times 10^{-4}$ and $1$ were chosen for the learning rate and $\lambda$, respectively. For data augmentation, we opted to only use image rotation by $180^{\circ}$ around the z-axis, which could simulate the supine and prone positions, after performing an investigation on the effects of more augmentation methods including image scaling. 

We used a five-fold cross validation to utilize all scans for evaluation. In addition to the global Dice and per case Dice scores used in the work by Tang et al.\cite{tang20}, we calculate per tumor Dice scores for all tumors and for relatively larger tumors ($\geq$ 125 mm$^3$, which is approximately $\geq$ 6 mm diameter) as well. While the per case Dice score denotes an average Dice score per scan/volume, the global Dice score is one computed by combining all CT scans into one. On the other hand, tight local image volumes around each tumor were taken into account to calculate the per tumor Dice scores. Also, paired t-tests are conducted to show the statistical significance of the proposed method.

\section{RESULTS}\label{sec:results}
%2p
\subsection{Quantitative Evaluation}

\begin{table}[t]
\centering
\caption{Quantitative comparison of different methods. `Seg' denotes performing only tumor segmentation from input CT volumes; `Seg + PP' denotes applying a post processing (PP) that is based on the tumor intensity likelihood to the results of `Seg'; `Seg + IL(in)' denotes using the generated likelihood volume as an additional input channel instead of as an additional output channel; `Seg + IL' denotes the proposed method; `Seg + IL(shifted)' denotes the proposed method but using a shifted likelihood function, which would be irrelevant with our target tumor. Dice scores were calculated at different subject levels, namely, global, per case, and per tumor. Refer to the text for the explanation on each of the metrics. Mean and standard deviation values are presented together, except for the global Dice scores. P-values are computed by conducting paired t-tests between the proposed method and the others with the Dice scores.}\label{tab:quan_res}
%\begin{scriptsize}
\begin{tabular}{c|c|c|c|c|c|c|c}
\multirow{2}{*}{Method} & Global & \multicolumn{2}{c|}{Per case} & \multicolumn{2}{c|}{Per tumor} & \multicolumn{2}{c}{Per tumor ($\geq$ 125 mm$^3$)} \\
\cline{2-8}
& Dice (\%) & Dice (\%) & p-value & Dice (\%) & p-value & Dice (\%) & p-value \\
\hhline{========}
Seg~\cite{cicek16} & 33.5 & 41.3 $\pm$ 27.2 & 0.0022 & 30.0 $\pm$ 36.7 & 0.0398 & 37.7 $\pm$ 36.4 & 0.1002 \\
\hline
Seg + PP & 27.8 & 36.2 $\pm$ 25.3 & 0.0001 & 24.7 $\pm$ 32.5 & 0.0005 & 30.5 $\pm$ 31.3 & 0.0026 \\
\hline
Seg + IL(in) & 31.4 & 41.6 $\pm$ 29.2 & 0.0860 & 32.8 $\pm$ 39.0 & 0.1885 & 37.9 $\pm$ 38.3 & 0.2133 \\
\hline
Seg + IL & \textbf{38.2} & \textbf{47.8} $\pm$ 29.6 & - & \textbf{35.9} $\pm$ 40.0 & - & \textbf{42.6} $\pm$ 39.5 & - \\
\hline
Seg + IL(shifted) & 29.8 & 41.1 $\pm$ 30.4 & 0.0084 & 30.1 $\pm$ 38.3 & 0.0296 & 32.4 $\pm$ 37.7 & 0.0025 \\
\end{tabular}
%\end{scriptsize}
\end{table}

Table~\ref{tab:quan_res} provides quantitative results of different methods, especially in terms of different ways of using the tumor intensity distribution information. Applying a post processing to the prediction of the tumor segmentation network, where the intensity likelihood volume $Y_{IL}$ is multiplied with the network predicted probability map, rather worsened the performance (`Seg + PP'). This may be because it could oversimply rule out tumors that have intensity values deviating from the built intensity model. We also tried using the intensity likelihood volume as an additional input channel instead of as an additional output channel (`Seg + IL(in)'). It can be another way to highlight our region of interest at input level. However, it performed on par with the baseline that does not use this additional input channel, `Seg'. On the other hand, the proposed method, `Seg + IL', showed clear improvements for all types of Dice score when compared to the baseline. We note that the proposed method of using intensity distribution supervision does not entail any additional labeling effort. It can be included in training by looking up already available CT scans and corresponding GT tumor segmentation to define the tumor intensity likelihood model.

We also investigate the effect of having the precise intensity model of targets. `Seg + IL(shifted)' follows the proposed method but uses another likelihood function that is $+100$ shifted from the original one. The shifted function does not reflect the actual intensity distribution of our target tumor any more. It performed rather worse than the baseline.  

%While carcinoid tumors in our dataset are overall small (the maximum tumor volume is 1960 mm$^3$, which is approximately 16 mm diameter), all methods showed higher Dice scores for relatively larger tumors ($\geq$ 125 mm$^3$, which is approximately $\geq$ 6 mm diameter) than for all tumors.
The Dice scores in Table~\ref{tab:quan_res} tend to be low since carcinoid tumors in our dataset are small (Figure~\ref{fig:stat}). Even small numbers of false positive and false negative voxels have a big impact on the Dice score of small lesions. Nevertheless, the proposed method showed clear improvements compared to the baseline. All methods including the proposed method showed higher Dice scores for relatively larger tumors ($\geq$ 125 mm$^3$, which is approximately $\geq$ 6 mm diameter) than for all tumors.

\subsection{Qualitative Evaluation}

\begin{figure}[t]
	\centering
	\begin{minipage}{1\textwidth}
        \subfloat{\includegraphics[width = 0.2\textwidth]{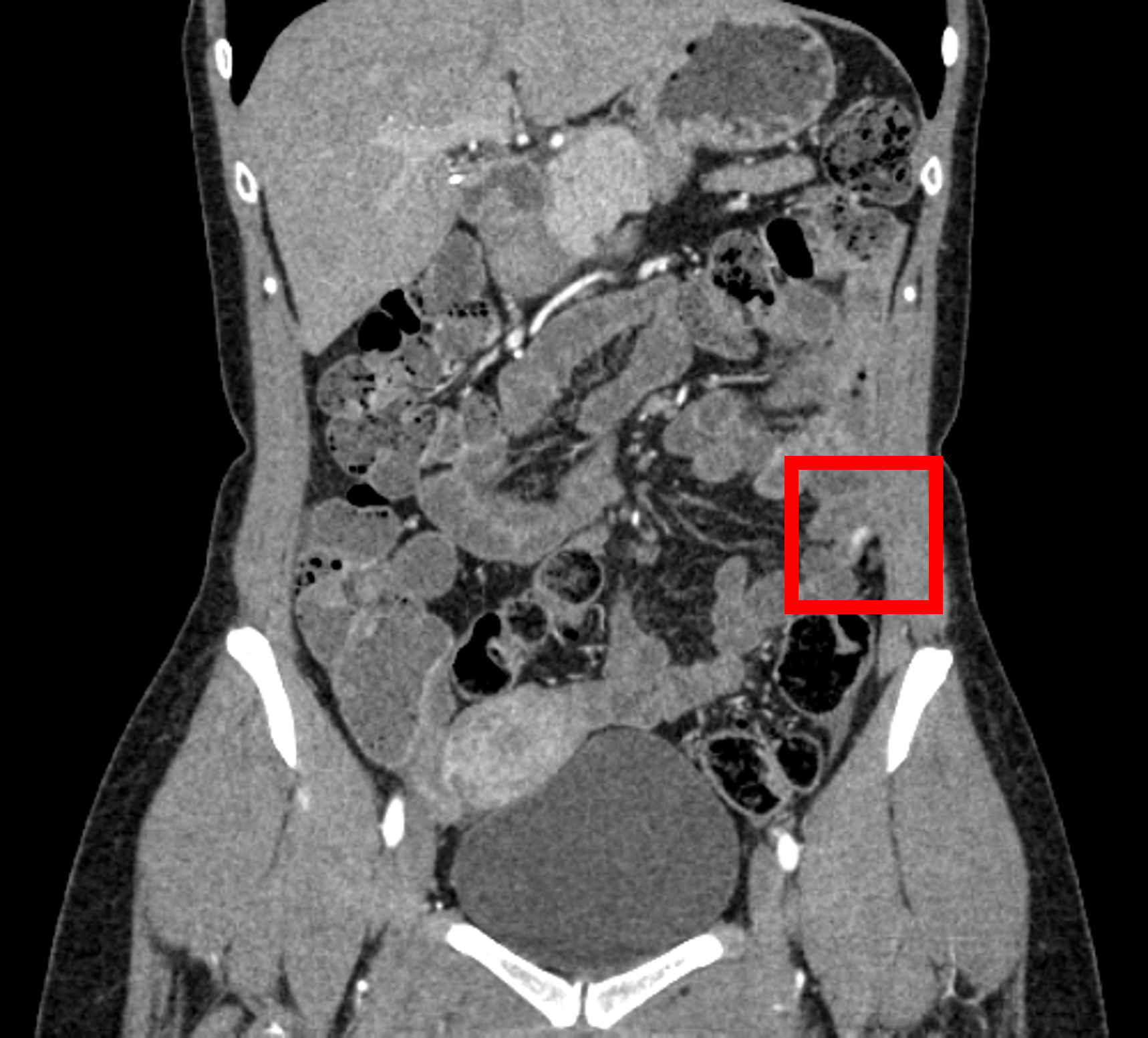}}
	    \subfloat{\includegraphics[width = 0.2\textwidth]{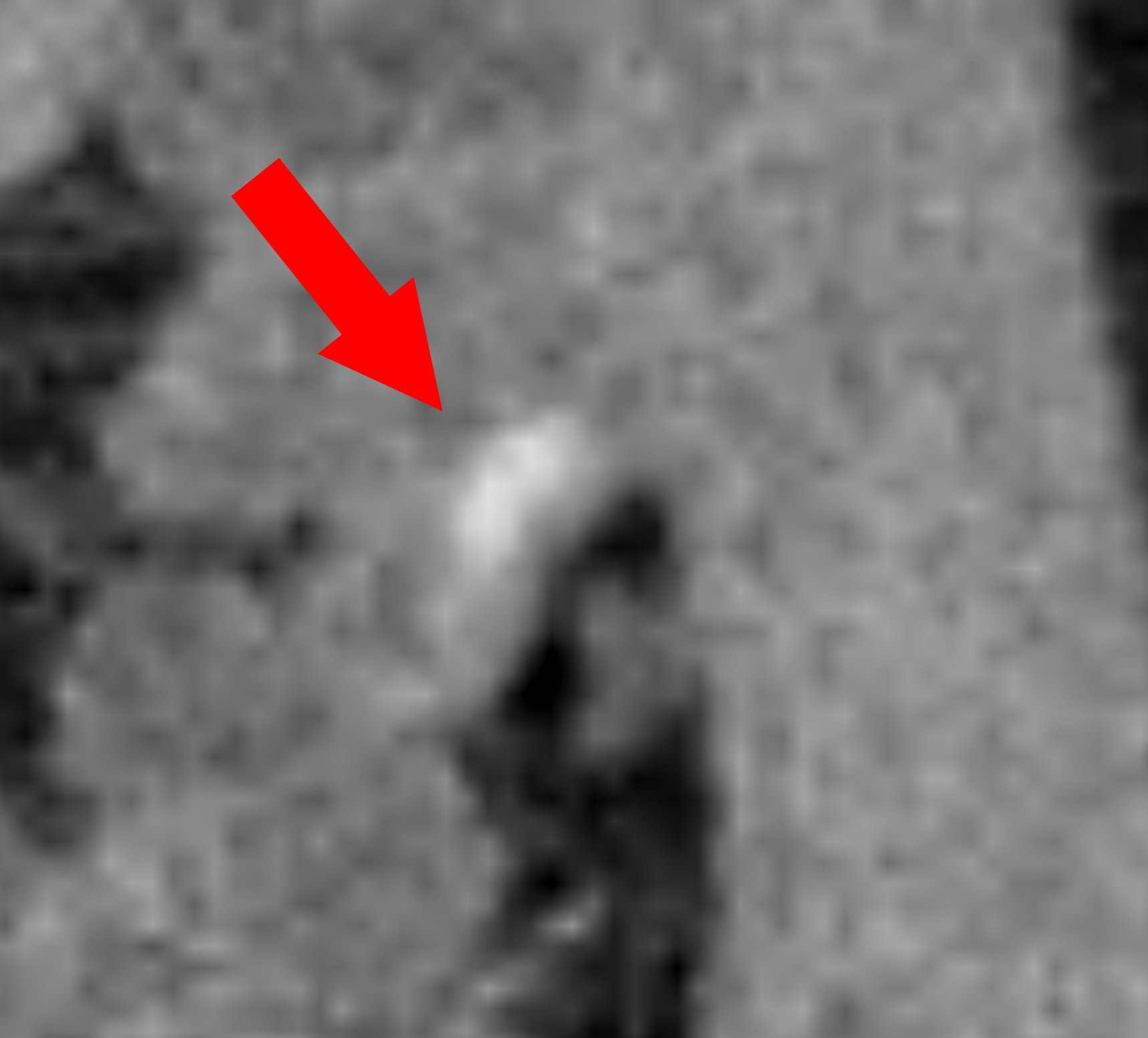}}
	    \subfloat{\includegraphics[width = 0.2\textwidth]{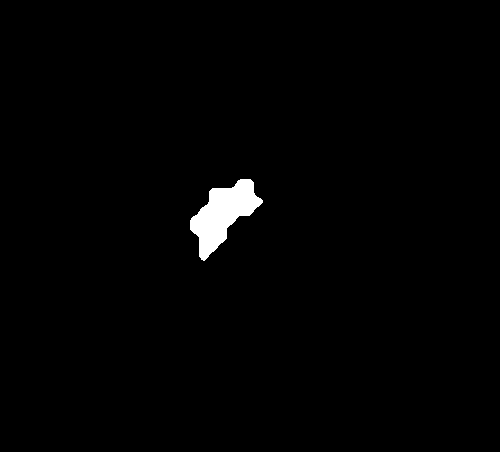}}
	    \subfloat{\includegraphics[width = 0.2\textwidth]{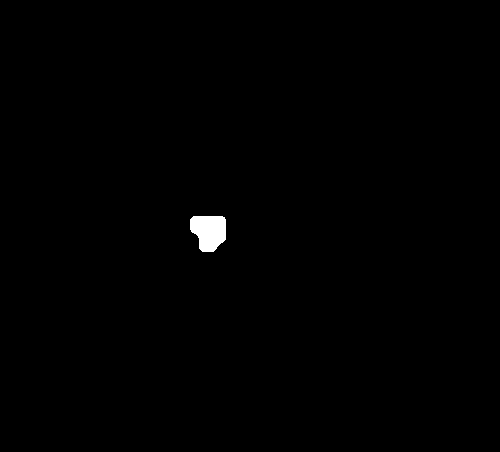}}
	    \subfloat{\includegraphics[width = 0.2\textwidth]{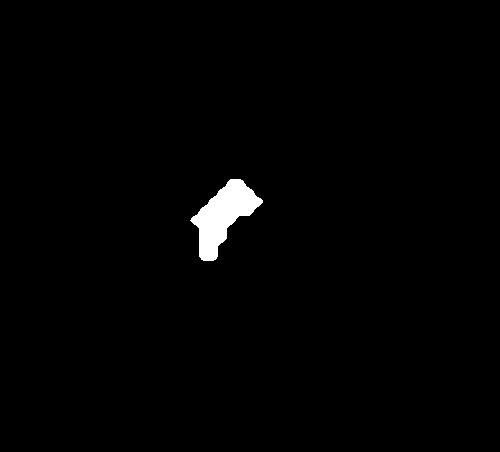}}
	    \vspace{-0.35cm}
    \end{minipage}
    \begin{minipage}{1\textwidth}
        \subfloat{\includegraphics[width = 0.2\textwidth]{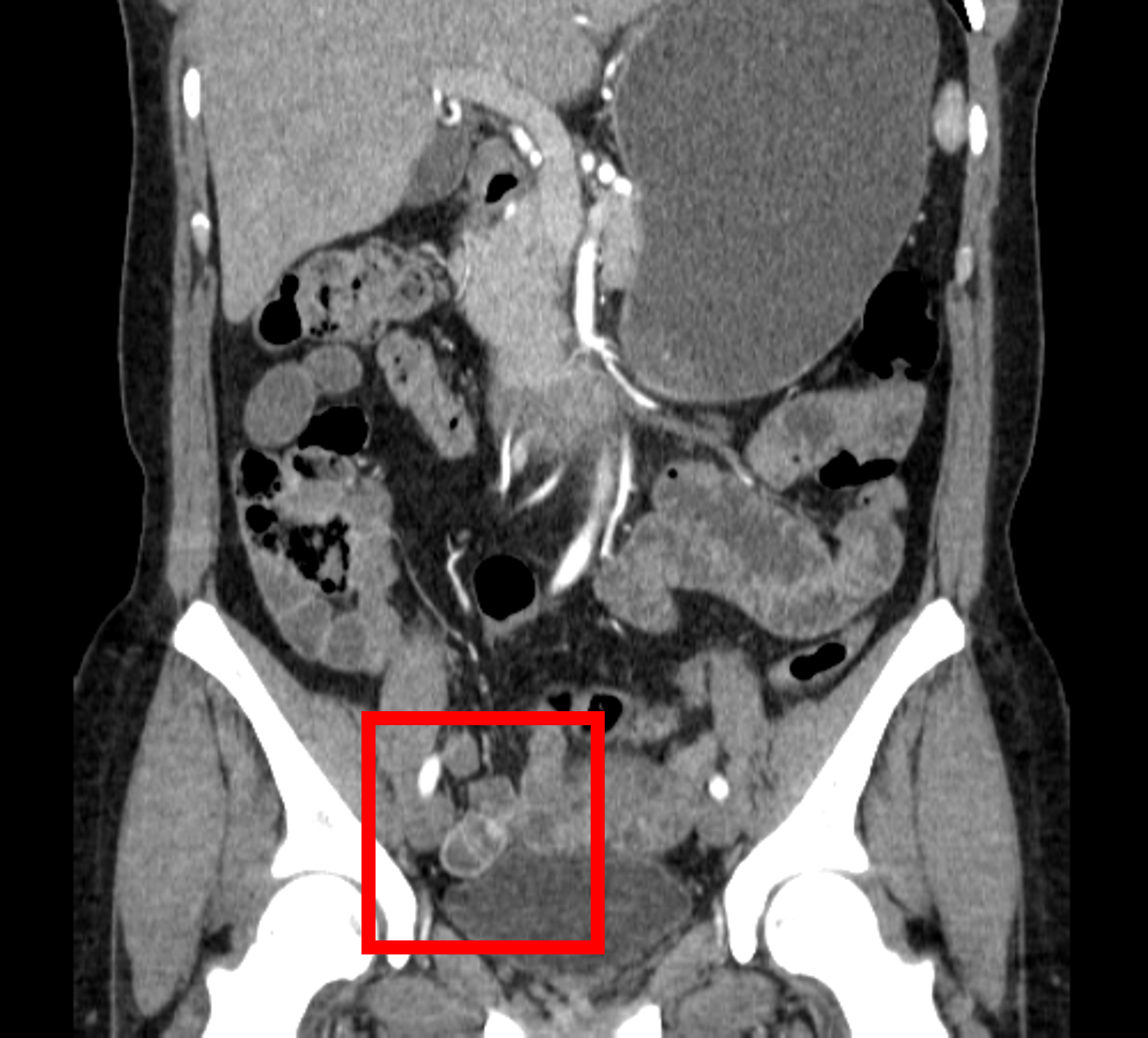}}
	    \subfloat{\includegraphics[width = 0.2\textwidth]{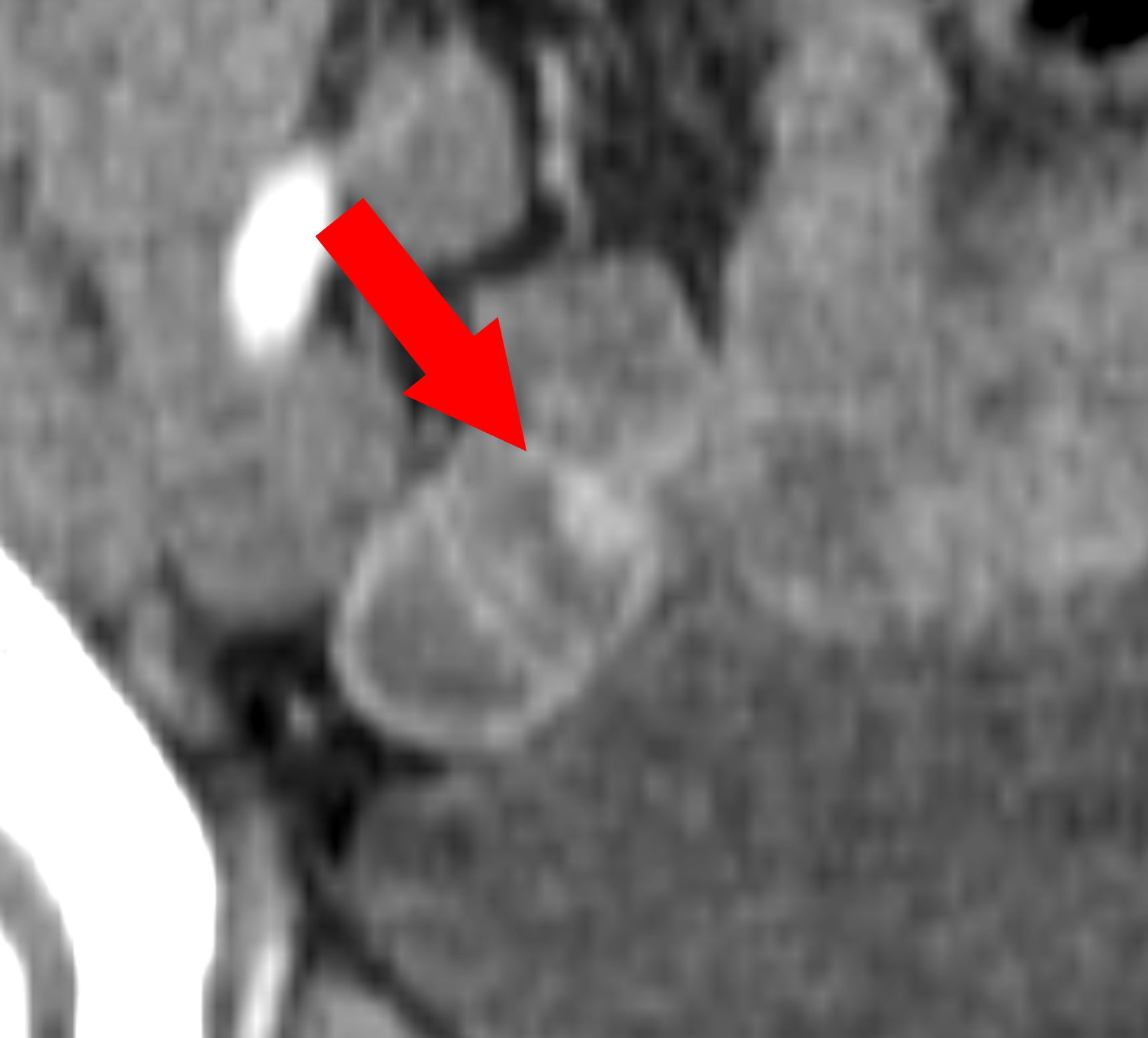}}
	    \subfloat{\includegraphics[width = 0.2\textwidth]{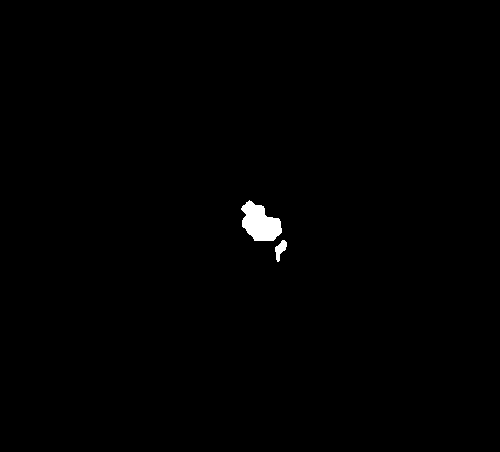}}
	    \subfloat{\includegraphics[width = 0.2\textwidth]{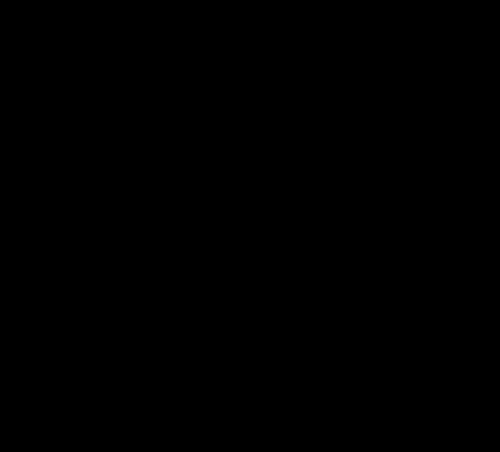}}
	    \subfloat{\includegraphics[width = 0.2\textwidth]{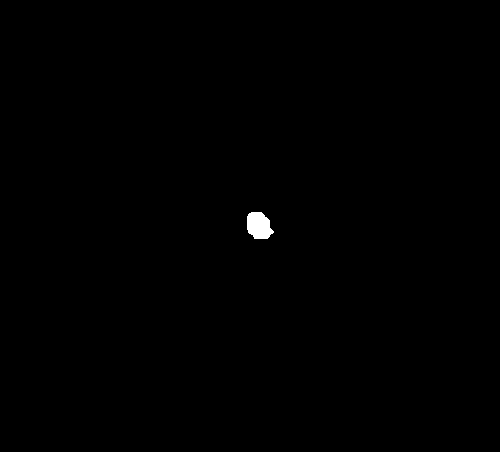}}
	    \vspace{-0.35cm}
    \end{minipage}
    \begin{minipage}{1\textwidth}
        \subfloat{\includegraphics[width = 0.2\textwidth]{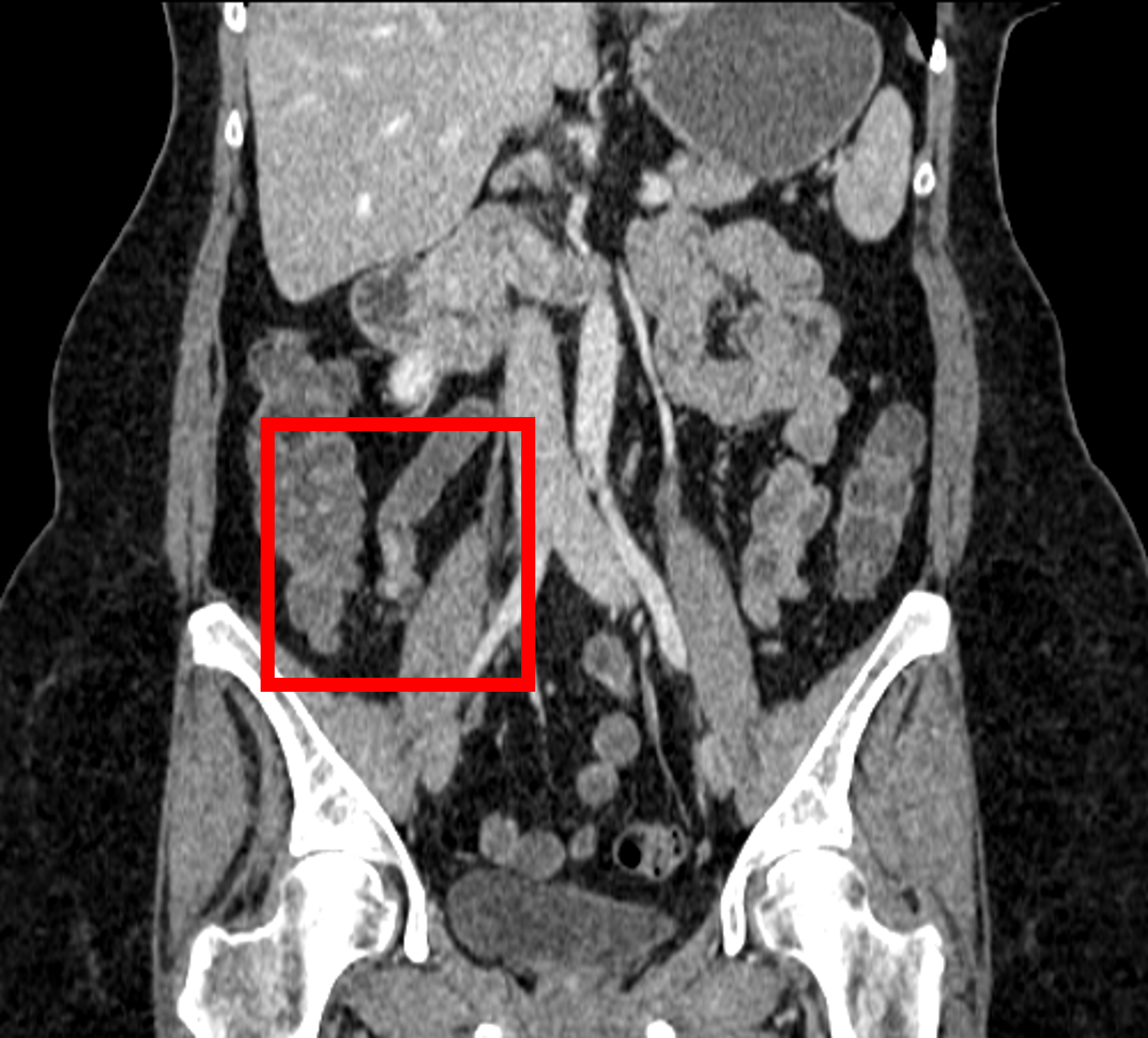}}
	    \subfloat{\includegraphics[width = 0.2\textwidth]{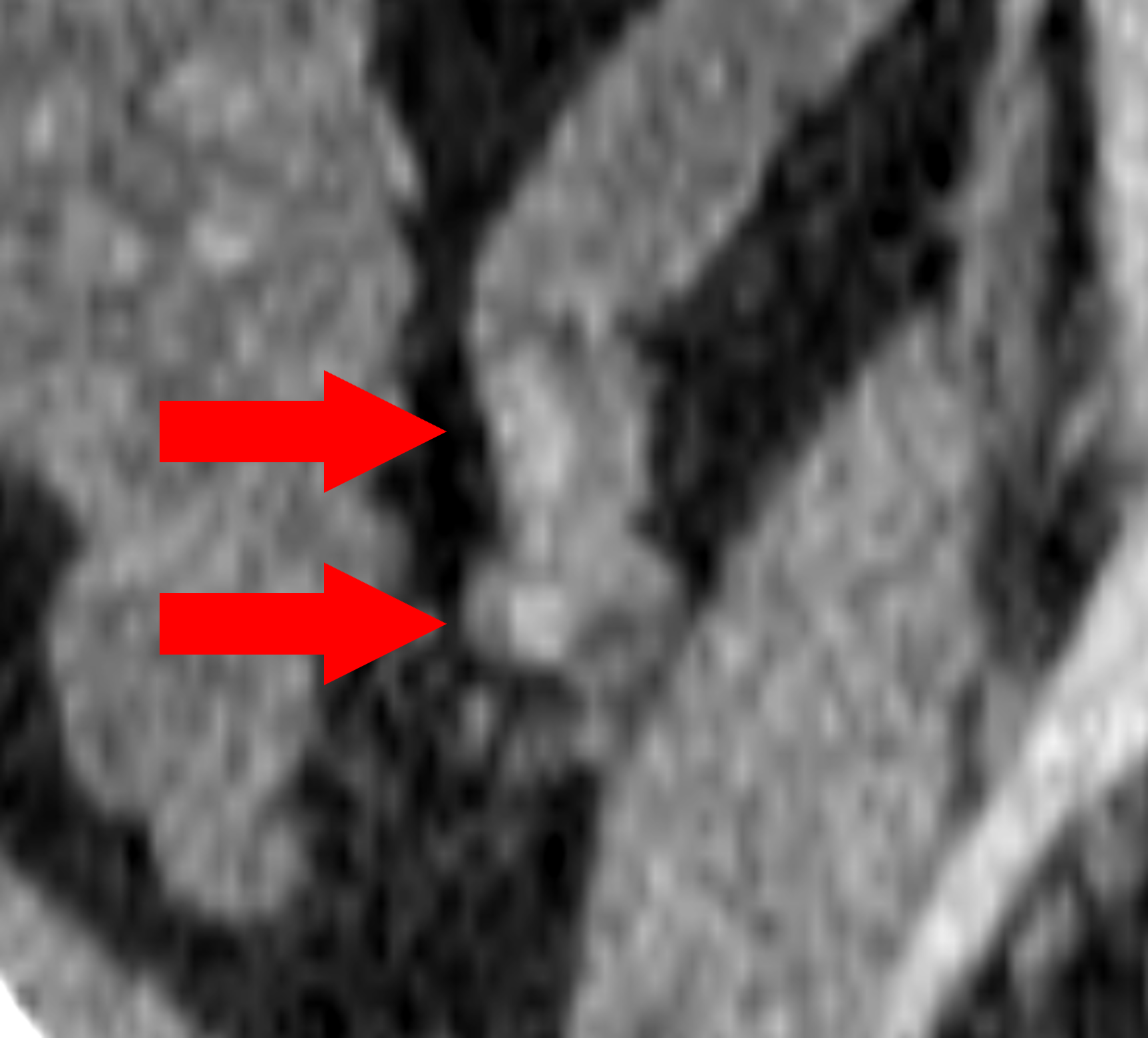}}
	    \subfloat{\includegraphics[width = 0.2\textwidth]{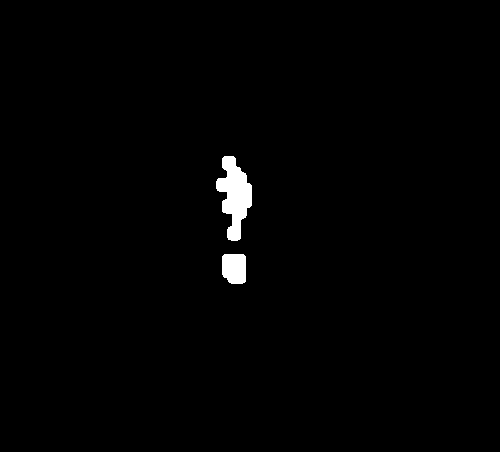}}
	    \subfloat{\includegraphics[width = 0.2\textwidth]{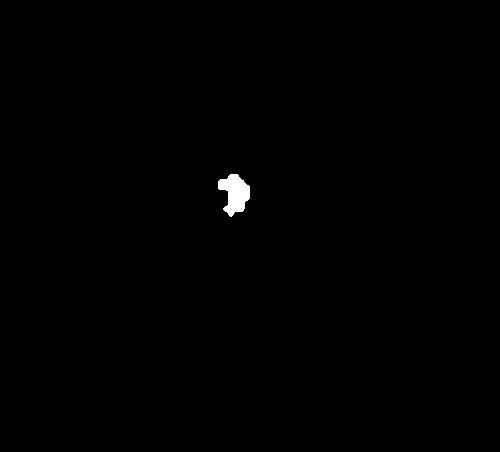}}
	    \subfloat{\includegraphics[width = 0.2\textwidth]{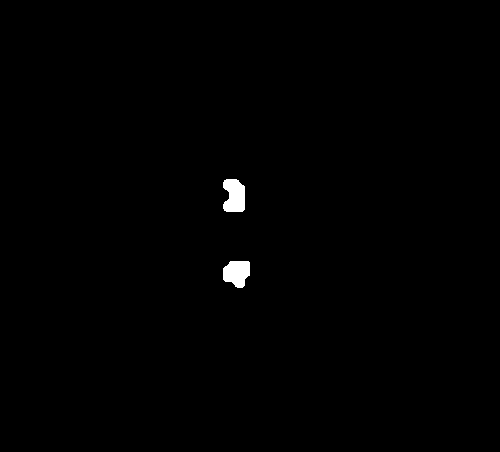}}
	    \vspace{-0.35cm}
    \end{minipage}
    \begin{minipage}{1\textwidth}
        \subfloat{\includegraphics[width = 0.2\textwidth]{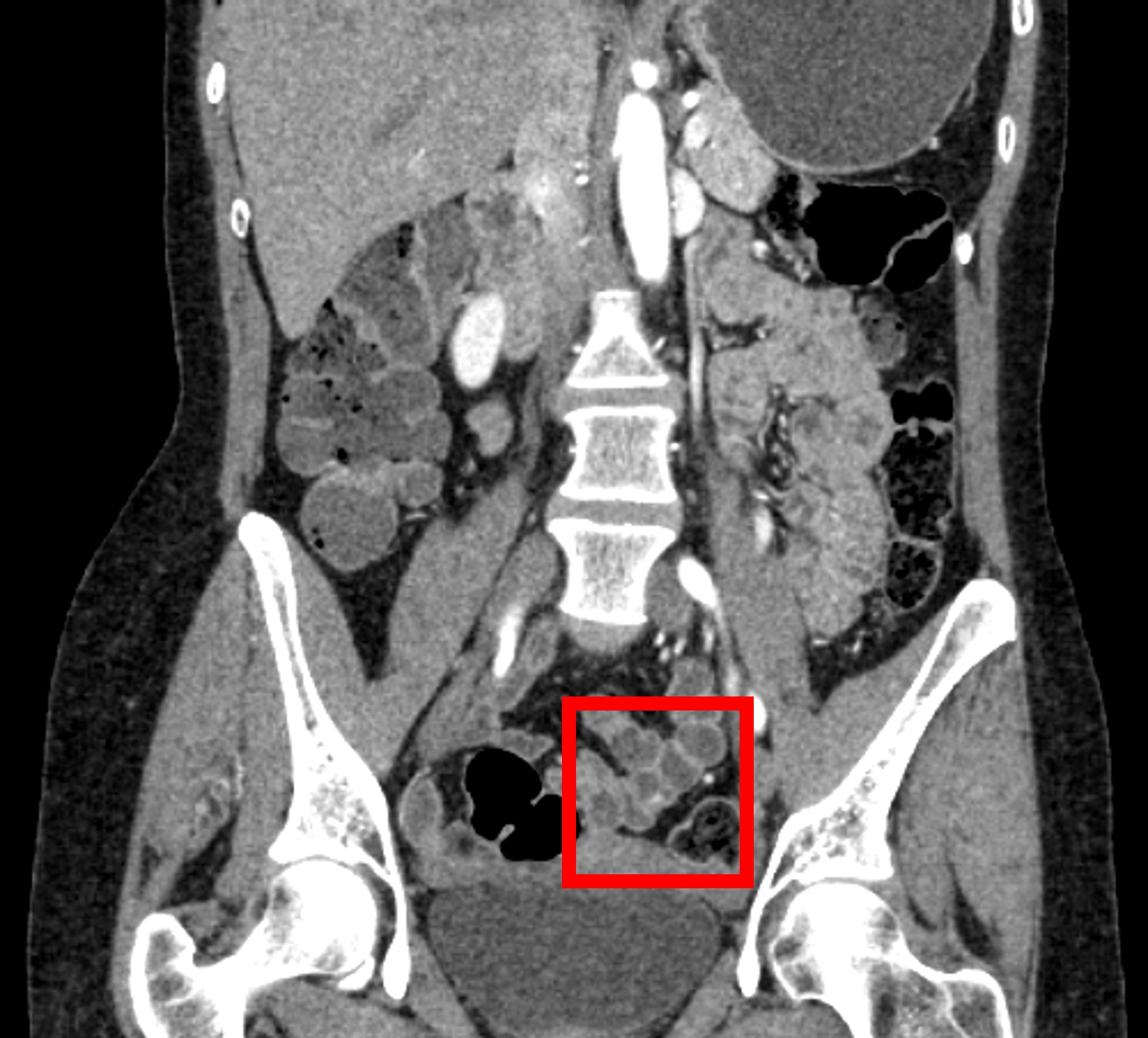}}
	    \subfloat{\includegraphics[width = 0.2\textwidth]{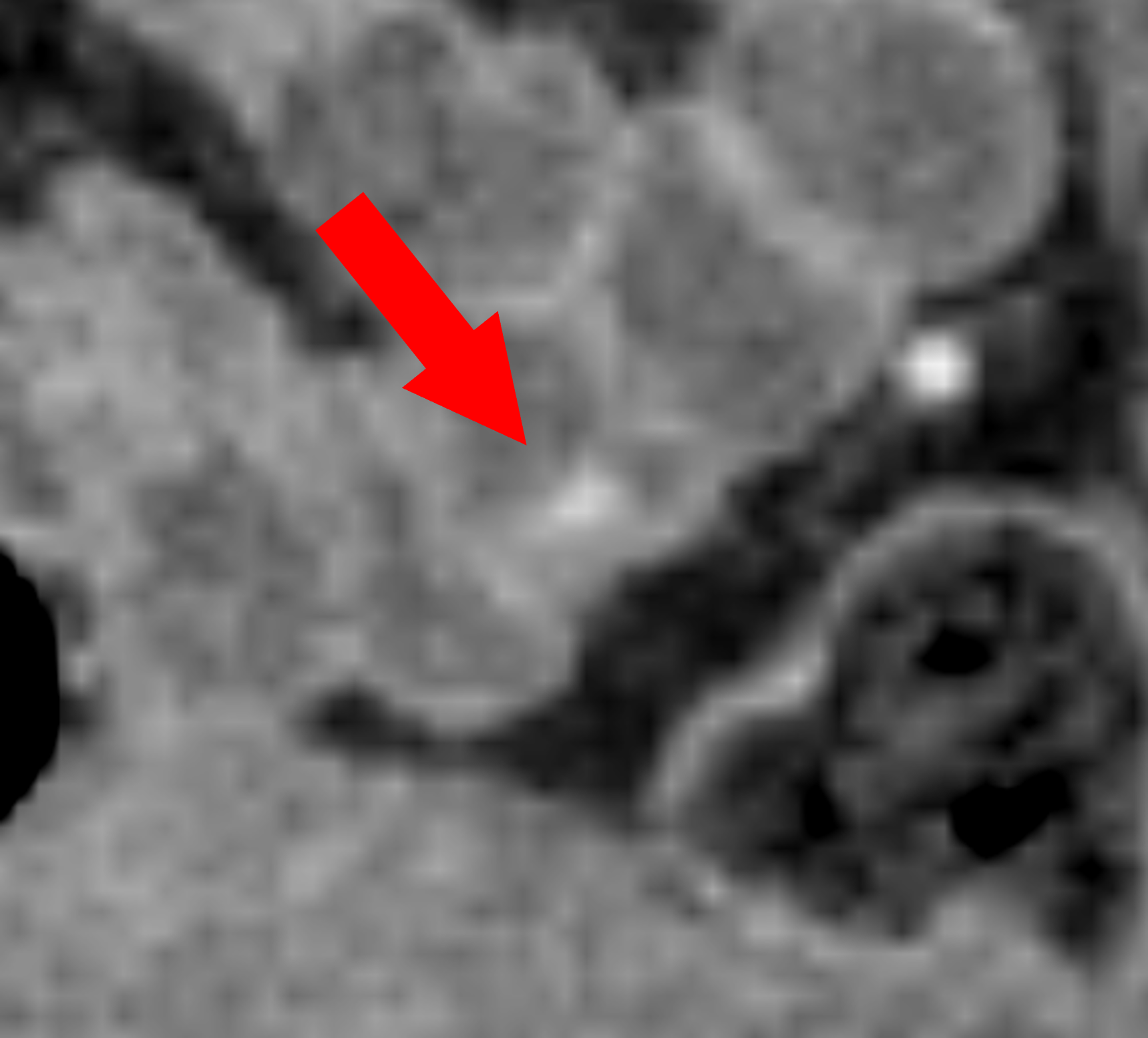}}
	    \subfloat{\includegraphics[width = 0.2\textwidth]{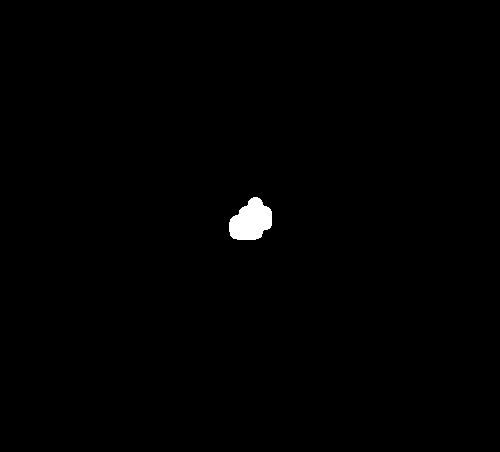}}
	    \subfloat{\includegraphics[width = 0.2\textwidth]{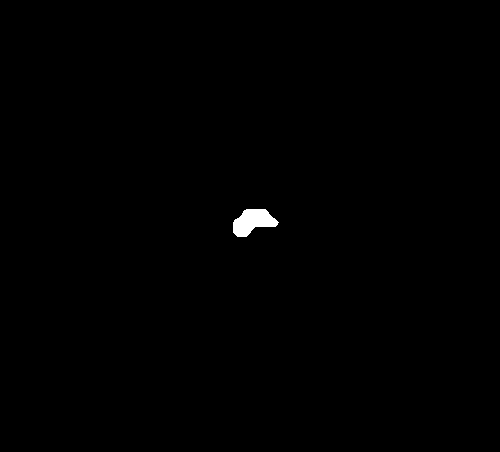}}
	    \subfloat{\includegraphics[width = 0.2\textwidth]{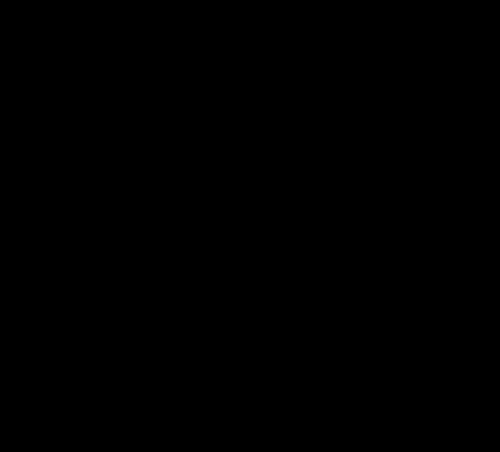}}
	    %\vspace{-0.35cm}
    \end{minipage}
	\caption{Example segmentation results of small bowel carcinoid tumor. Each row represents different cases. The columns, from left, represent input CT scan, zoomed view of the red box in the CT scan (The tumors are pointed by the red arrow), corresponding GT tumor segmentation, result of the baseline method (`Seg' in Table~\ref{tab:quan_res}), and result of the proposed method (`Seg + IL' in Table~\ref{tab:quan_res}), respectively.}
	\label{fig:qual_res}
\end{figure}

\begin{figure}[t]
	\centering
	\begin{minipage}{1\textwidth}
        \subfloat[]{\includegraphics[width = 0.25\textwidth, frame]{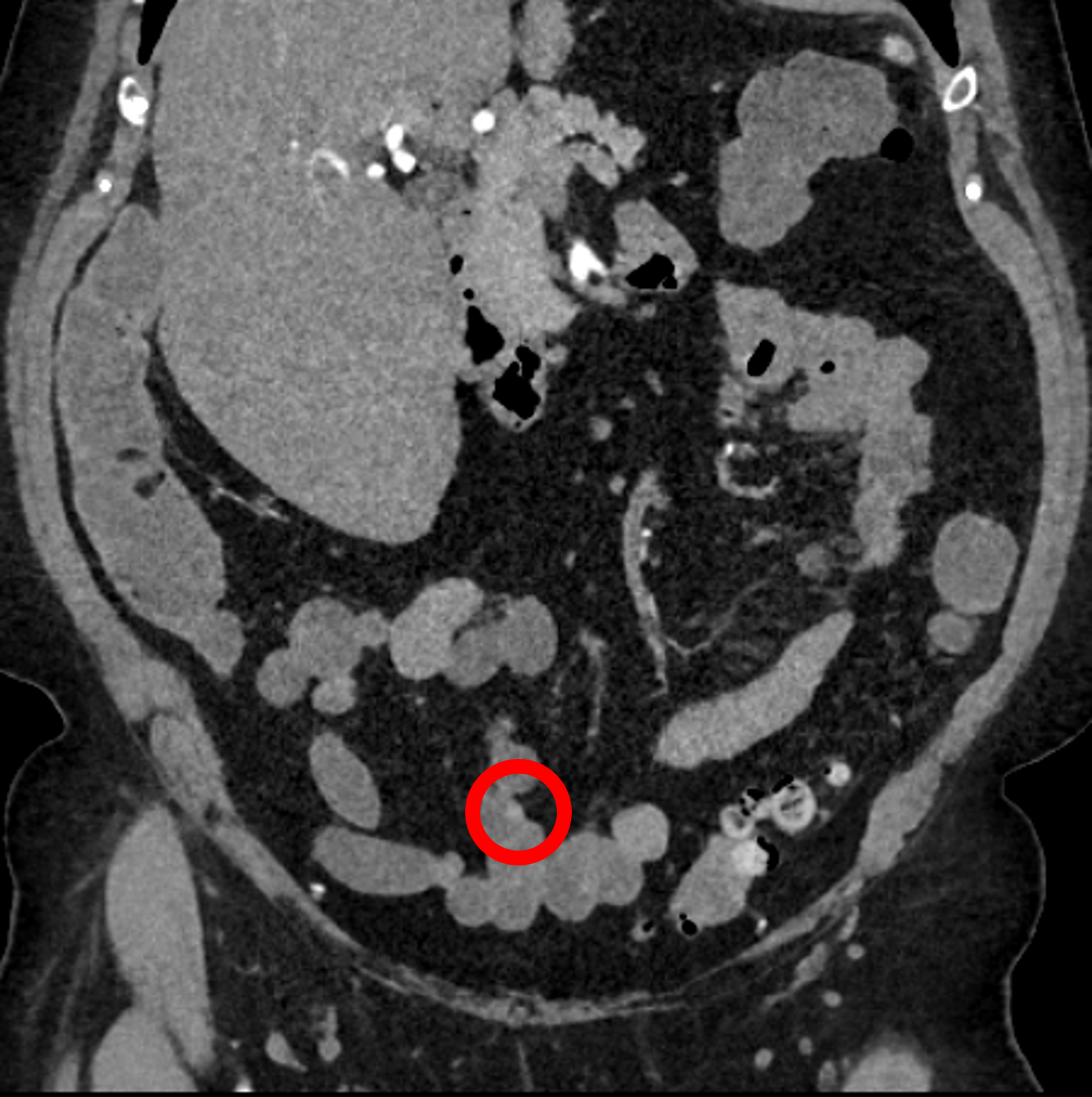}}
	    \subfloat[]{\includegraphics[width = 0.25\textwidth, frame]{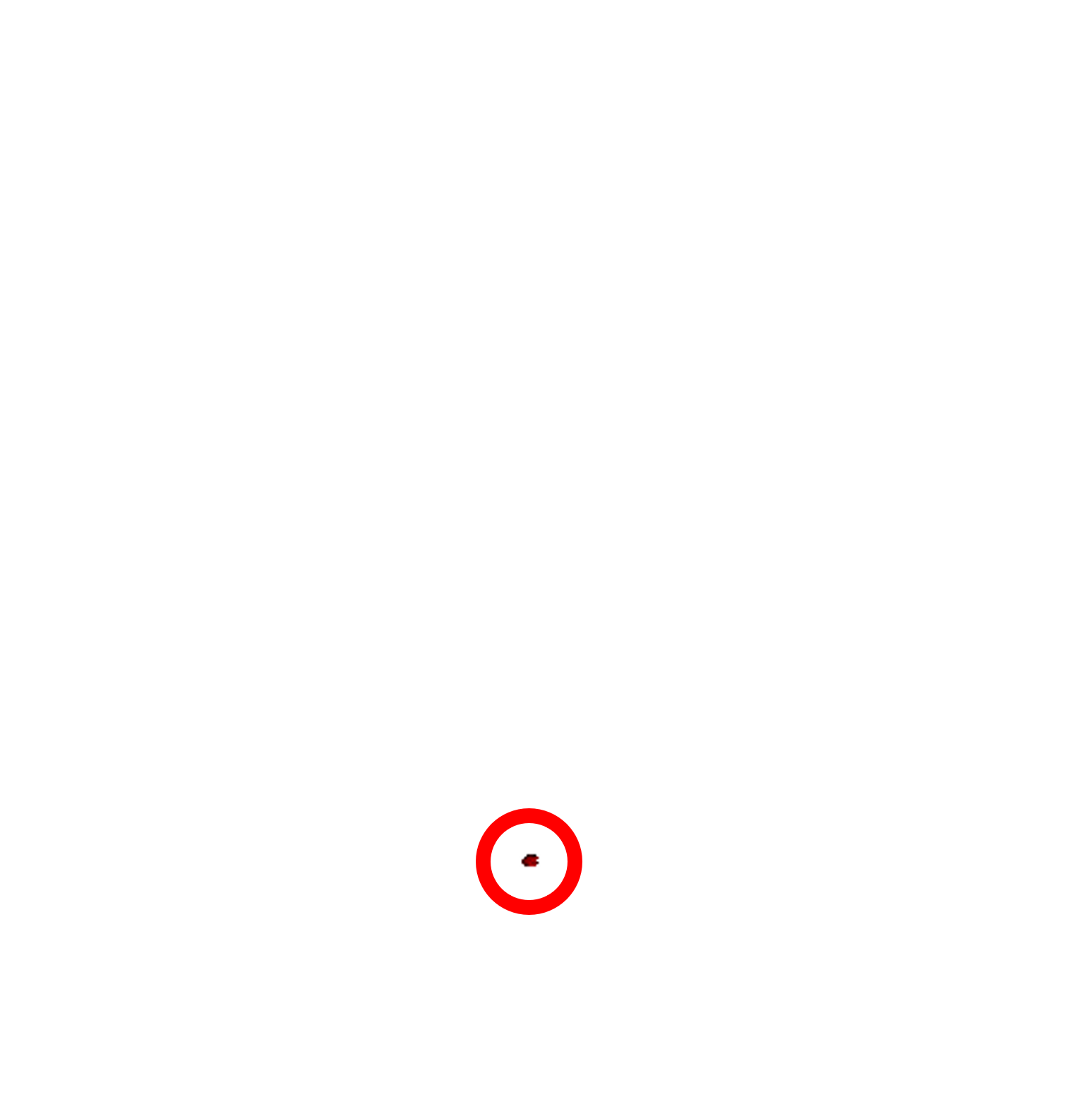}}
	    \subfloat[]{\includegraphics[width = 0.25\textwidth, frame]{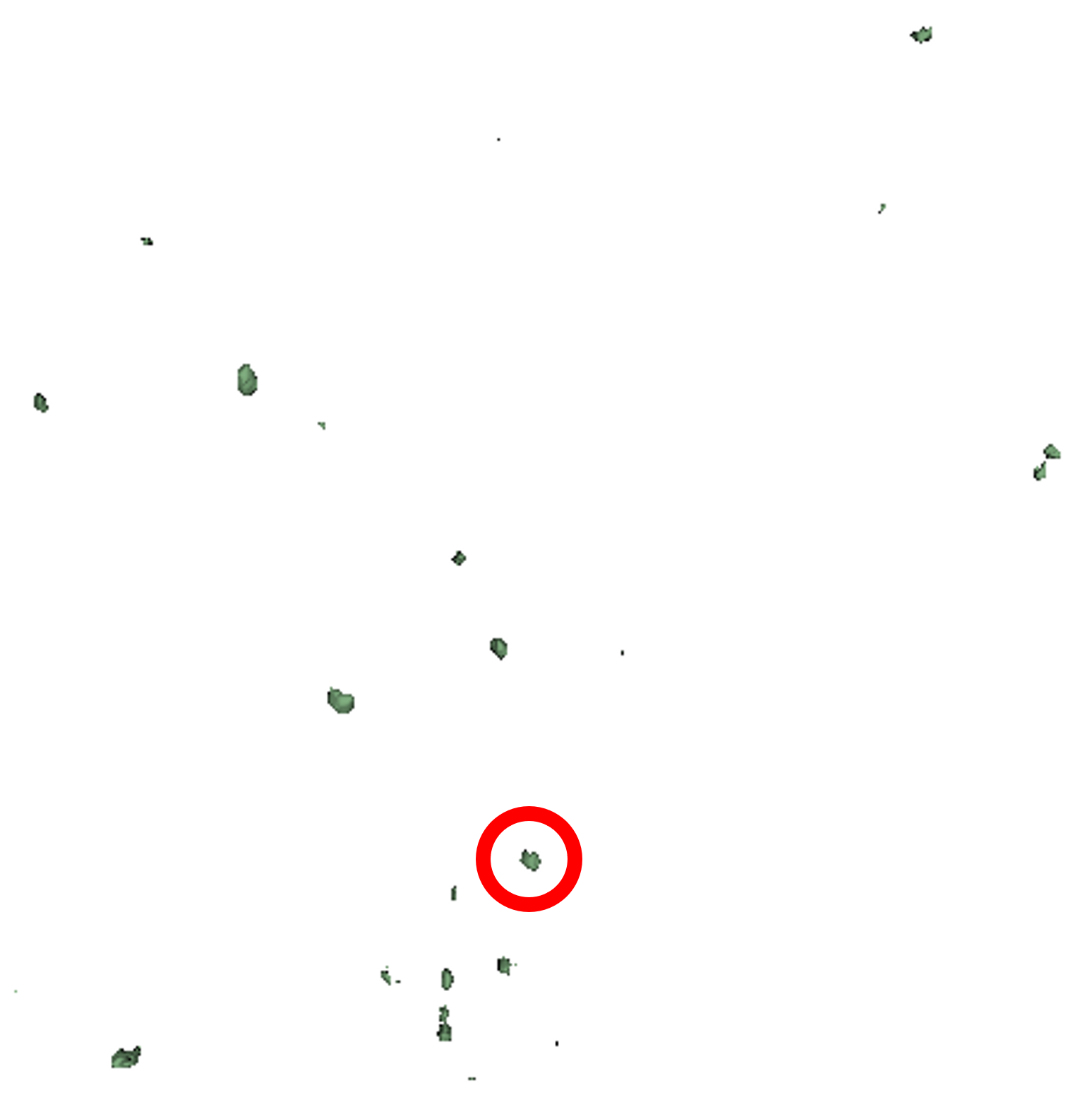}}
	    \subfloat[]{\includegraphics[width = 0.25\textwidth, frame]{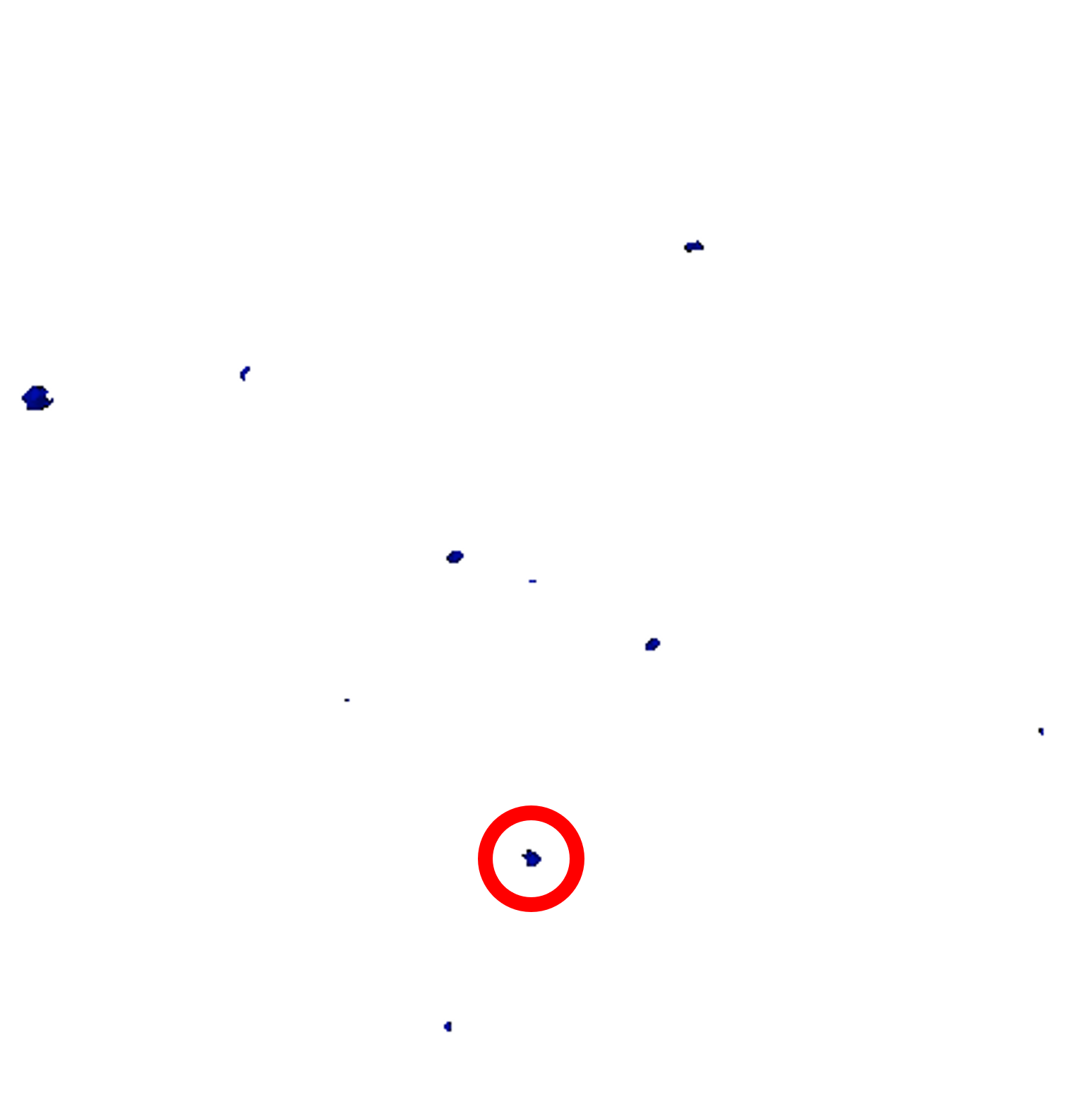}}
    \end{minipage}
	\caption{Example segmentation results in 3D. (a) Input CT scan that has one small bowel carcinoid tumor, which is highlighted by the red circle. An image slice in coronal view is shown. (b) GT segmentation. (c) Result of the baseline method (`Seg' in Table~\ref{tab:quan_res}). (d) Result of the proposed method (`Seg + IL' in Table~\ref{tab:quan_res}).}
	\label{fig:qual_res_3d}
\end{figure}

Figure~\ref{fig:qual_res} shows example segmentation results of small bowel carcinoid tumor. Compared to the baseline method that is trained without the intensity distribution supervision, the proposed method segments the entire tumor area better (first row), or segments more tumors (second and third rows). The last row represents a failure case, where the proposed method missed a blurry small tumor. Figure~\ref{fig:qual_res_3d} presents an example result in 3D. While both of the baseline and the proposed methods segment the GT tumor successfully, the proposed method has less false positive segmentation.

\section{CONCLUSION}\label{sec:conclusion}
%0.3p
We have presented a method for improving small lesion segmentation in CT scans, which utilizes intensity distribution of the target lesion. It requires no additional labeling effort. We applied the proposed method to segmentation of small bowel carcinoid tumors, and the experimental results proved the validity of our idea. The improved method could better assist a radiologist with the challenging task of finding carcinoid tumors in the small bowel on CT scans. In future work, we plan to apply the method to different target lesions to further validate the efficacy of the proposed method.

\acknowledgments % equivalent to \section*{ACKNOWLEDGMENTS}       
 
This research was supported by the Intramural Research Program of the National Institutes of Health, Clinical Center. The research used the high performance computing facilities of the NIH Biowulf cluster. 
 
% References
%\bibliography{ref} % bibliography data in report.bib
%\bibliographystyle{spiebib} % makes bibtex use spiebib.bst

\end{document}